\documentclass[aps,prb,showpacs]{revtex4}
\usepackage{graphicx}

\begin{document}
\title{Strong two-photon emission by a medium with periodically
time-dependent refractive index}

\author{V. Hizhnyakov}
\affiliation{
Institute of Physics, University of Tartu, Riia 142\\
51014 Tartu, Estonia\\
e-mail: hizh@fi.tartu.ee\\}


\begin{abstract}

A two-photon emission of a medium with periodically time-dependent
refractive index is considered. The emission results from the
zero-point fluctuations of the  medium. Usually this emission is
very weak. However, it can be strongly enhanced if the resonant
condition $\omega_0$ = 2.94 $c/l_0$ is fulfilled (here $\omega_0$
and $l_0 $ are the frequency and the amplitude of the oscillations
of the optical length of the medium, respectively). Besides, a
medium with resonant oscillations of the optical length performs
the phase conjugated reflection with high efficiency. A similar
resonant enhancement of the two-quantum emission of other bosons
is also predicted.

\end{abstract}

\pacs{PACS numbers: 42.50. Wm,\,\, 03.08.+r,\,\, 42.65.Re}
\maketitle

The usual nonlinear optical processes leading to the generation of
new waves, account for the stimulated emission \cite{Shen}.
Unlikely to that, the two-photon emission of a medium with the
refractive index changing in time occurs due to the spontaneous
emission arising from the zero-point fluctuations of the field
\cite{Casimir,Jablonovich,quantopt}. An interesting heuristic
aspect of this emission is its close relation to some basic
processes in quantum cosmology, e.g. to the creation of matter in
the initial stage of the expanding universe \cite{quantopt}.
Besides, the change of $n$ linear in time results in the
thermal-like radiation \cite{Jablonovich} "seen" by an observer
when it moves with the constant acceleration \cite{Unruh}; this
radiation is closely related to the Hawking radiation of black
hole \cite{Birel}.

As a rule, the two-photon emission under consideration is very
weak. One can expect it to be observed only if the change of $n$
in time takes place in a large area and is very large and fast
\cite{quantopt}. The linear and strong time dependence of $n$ can
be realized only in a small "spot" in the medium for a very short
time \cite{Jablonovich}. Unlikely to that, the periodic and rather
substantial oscillations of $n$ in time can be activated in a
medium of a large size by applying a strong (quasi)monochromatic
laser beam.

The size factor is of a primary importance here. Indeed, the
oscillations of the refractive index in time lead also to
oscillations of the optical length of the medium $l$. Taking $n(t)
= n_0 + n'_0 \cos{\omega_0 t}$, we get for the time-dependent part
of the optical length $l(t)=l_0 \cos{\omega_0 t}$, where $l_0 =
n'_0 L^{(0)}$, $\,L^{(0)}$ is the length  of the coherently
excited medium.  Consequently, if  $L^{(0)}$ is large then one can
achieve large amplitude of the oscillations of the optical length
$l_0$ even if $n'_0$ is not too large. Therefore, the maximal
velocity of the oscillations of the optical length $V=\omega_0
l_0$  can also get large values which may be comparable, or even
exceed the velocity of light in vacuum $c$. This has an important
consequence: strong enhancement of the quantum emission if the
maximal velocity $V$ of the oscillations of the optical length
approaches $2.94 c$. The reason for this resonant enhancement can
be elucidated as follows. If $V \ll c$ then the intensity of the
emission increases with $V$. However, if $V \gg c$, then the
opposite dependence takes place, as the zero-point fluctuations
cannot appreciably react on such a fast oscillations. Therefore,
the crossover for the dependence of the emission on $V$ exists at
$V_r \sim c$ and a resonant enhancement of it is observed at $
V_r$. A possibility of the enhancement of the quantum emission in
a dielectric medium with time-dependent parameters has been
pointed out by a number of authors (see, e.g.
\cite{Lobashov,Klimov,Johnston,Saito}). But, in all previous
studies this enhancement has been found to take place only in
short resonator, which has discrete modes with well-separated
frequencies; the enhancement is observed only for the mode, the
multiple frequency of which is in resonance with $\omega_0$. For a
given resonator there are series of modes and resonances; their
frequencies are determined by the geometry of the resonator. The
emission is non-stationary: the number of generated photons
increases in time exponentially. However in our case the
enhancement takes place only at $\omega_0 = 2.94 c/ n'_0 L_0 $;
the emission is stationary. Such a big difference of the
properties of these phenomena results from the different
mechanisms of the enhancement: in the case of a short resonator
this is the parametric resonance for a time-dependent
electromagnetic mode. The stimulated emission plays here an
essential role: all generated photons remain in the resonator and
essentially exert the process, resulting in the exponential
increase of the number of photons in time. However in the case
under consideration there is a continuum of modes which are mixed
due to oscillations of $n$. The enhancement of the emission at $V=
2.94 c$ is due to the dynamical resonance of the zero-point
fluctuations with the oscillations of the optical length. All
generated photons immediately leave the region of generation and
do not influence the process. Therefore, only a spontaneous
emission contributes and the emission is stationary.

We are studying a medium exposed in a monochromatic standing laser
wave. We describe the wave classically and account for the
time-dependent part of the nonlinear polarization operator
$\hat{P}^{(nl)}$. The equation for the field operator $\hat{A}$
then reads \cite{Shen}
$\ddot{\hat{A}}- (c/n_0)^2 \hat{\nabla}^2 \hat{A} =- 4\pi
\ddot{\hat{P}}^{(nl)},$ where $\hat{P}^{(nl)}$ is the
time-dependent part of the nonlinear polarization operator.
We take $\hat{P}^{(nl)} (t) =\eta (t)\hat{A}/ 4 \pi$, where
$\eta(t) = \eta_{\,0} \,\cos{\omega_0 t}$. Then we get
\begin{equation}
\ddot{\hat{ A}}'- (c/n(t))^2 \hat{\nabla}^2 \hat{A}'=0,
\label{difeq}
\end{equation}
where $\hat{ A}'= \hat{A}(1+ \eta (t)),$ $\, n(t) = n_0\sqrt{1
+\eta(t)}$. We suppose that $\eta_0 \ll 1$; then $n(t) \approx n_0
+ n'_0 \cos{\omega_0 t}$, where $n'_0 \approx \eta_0/2n_0$.
Equation (\ref{difeq}) is the wave equation with a harmonically
time-dependent refractive index.

We are considering a dielectric situated in a very large resonator
and suppose that the refractive index of the dielectric changes in
time only in the time interval between $t=0 $ and $t=t_0 $. We
want to find how this change influences upon the vacuum state  of
the quantum field at $t > t_0$. To this end, we use the Coleman
theorem \cite{Coleman}, which asserts that the time-dependent
classical field leads to the change of the vacuum state in time.
In our case this testify that the initial ($\hat{a}_k ,\,
\hat{a}^+_k$) and final ($\hat{b}_k,\, \hat{b}^+_k$) destruction
and creation operators of a mode $k$ in the resonator are
different \cite{Birel}: they are related to each other by the
Bogoljubov transformation
\begin{equation}
\hat{b}_{k} = \mu_k\,\hat{a}_k + \nu_k\, \hat{a}^+_k,
\label{transf}
\end{equation}
where $|\mu_{k,q}|^2= 1+|\nu_{k,q}|^2.$  This means that photons
appear in the final state. The number of generated photons of the
mode $k$ equals $ N_k=\langle 0|\hat{b}^+_k \,(t)\, \hat{b}_k
\,(t) |0\rangle = |\nu_k|^2, $ where $|0\rangle$ is the initial
zero-point state (in this state $\hat{a}_k|0\rangle =0$); the
photons appear by pairs.

Usually, to find $N_k$, one calculates the parameters of the
Bogoljubov transformation (\ref{transf}) (see, e.g. \cite{Birel}).
However, there also exists another, a simpler way based on
calculation of the pair correlation function
\begin{equation}
 {\cal D}_k(t; \tau)= \langle 0| \hat{A}_k (t + \tau ) \hat{A}_k
(t)|0 \rangle  \label{calD}
\end{equation}
at a large time $t > t_0$, $\, \tau \ll t$ and with $t$ averaged
over a period of oscillations \cite{euro}; here $ \hat{A}_k (t) =
(\hbar/2\omega_k)^{1/2} \big(\hat{b}_k e^{- i \omega_k t} +
\hat{b}^+_k e^{i \omega_k t}\big)\, $, $\omega_k$ is the frequency
of the mode $k$. Indeed, inserting  Eq. (\ref{transf}) into Eq.
(\ref{calD}) we find
\begin{equation}
{\cal D}_{k}(t;\tau) = (\hbar/2\omega_{k}) \big(|\mu_{k}|^2
e^{-i\omega_{k} \tau }+ |\nu_{k}|^2 e^{i\omega_{k} \tau }\big),
\label{Dasymp}
\end{equation}
(the terms $\propto e^{\pm 2i\omega_k t}$ drop out). Consequently,
to find the number of the generated photons $N_k$, one may
calculate the negative frequency (with respect to $\tau$) term of
the large-time $t$ asymptotic of the pair correlation function
${\cal D}_{k}(t;\tau)$. Below we will use this method to calculate
the quantum emission under consideration.

We suppose that $n'_0$ very slowly changes in space. Then,
according to the theory of the wave-optics \cite{Solimeno} the
plane waves $\hat{A}_q (x,t) \propto e^{i q x}$ are the solutions
of this equation outside the medium (then the operators $\hat{A}'$
and $\hat{A}$ coincide); at that the allowed values of the wave
number $q$ satisfy the condition $q L(t) = 2 \pi k$, where $k =
0,1,2 \dots,$  $L(t)= L_0 + l(t)$ is the optical length (eikonal)
of the resonator + dielectric at time $t$. Therefore the field
operator outside the medium has the form
$$
\hat{A}(x,t) = \sum_k    \sin{(\pi k x / L(t))} \hat{A}_{k}(t).
\label{A1}.
$$
This  operator satisfies the wave equation in vacuum
$\ddot{\hat{A}} - c^2 \nabla^2 \hat{A} =0$ which gives in the
$L/l_0 \rightarrow \infty$ limit
\begin{equation}
\sum_{k} \bigg[\ddot{\hat{A}} _{k} + \omega^2_{k} \hat{A}_{k} -
\frac{k \pi x }{L^2}\big(2\dot{L} \dot{\hat{A}}_{k,{\vec q}}  +
\ddot{L} \hat{A}_{k}\big) \cot{\bigg(\frac{\pi k x}{L}\bigg)}
\bigg] =0,  \label{waveeq}
\end{equation}
where $\omega_{k} = \pi c k /L_0$ (the terms $ \propto L^{-m}$
with $m > 2$ are neglected). Taking into account the identity
$$
\pi x/L =- 2 \sum_{j=1}^{\infty} (-1)^j j^{-1} \sin{(\pi j x /L)},
\quad -L \leq x \leq L,
$$
one gets for $\hat{{\cal A}}_k = (-1)^k \hat{A}_k$\, the equation
%
$\ddot{\hat{\cal A}}_{k} + \omega^2_{k} \hat{\cal A}_{k} \simeq
\omega_k \hat{B}_{k},$ where
$$ \hat{B}_k=  (2/\pi c) \sum_{j \neq k}j
(2 \dot{L} \,\dot{\hat{{\cal A}}}_j + \ddot{L} \, \hat{{\cal
A}}_j)/(j^2-k^2). $$

We consider the case when  the  oscillations of $n$ last for a
long time $t_0$. Using the Green function of the harmonic
oscillator $\omega_k^{-1} \Theta{(t)} \sin{\omega_k t}$, where
$\Theta{(t)}$ is the Heaviside step-function, we get for $t \geq
t_0$
\begin{equation}
\hat{{\cal A}}_k(t) \simeq \hat{{\cal A}}^{(0)}_k (t) +
\int_0^{t_0} \!\!\! dt_1 \sin{(\omega_k(t-t_1))} \! \hat{B}_k
(t_1), \label{intA}
\end{equation}
where $\hat{{\cal A}}^{(0)}_k (t)= (-1)^k (\hbar/2 \omega_k) (
\hat{a}_{k} e^{-i \omega_k t}+ \hat{a}^+_{k}e^{i \omega_k t})$.
From Eq. (\ref{intA}) it follows that in the $t \rightarrow
\infty$ limit the operator $\hat{\cal A}_k(t)$ consists of two
items: the positive frequency item $\propto e^{-i\omega_{k}\,t}$
and the negative frequency item $\propto
e^{i\omega_{k,\vec{q}}\,t}$. Besides,  in the large $t_0$ limit
the main contribution to the integral (\ref{intA}) comes from
large $t_1$. In this case the time-dependence of the factors
$\hat{{\cal A}}_{j}$ entering the equation for $\hat{B_k}$, is
given by the exponents $e^{\pm i\omega_{k}\,t_1}$. We also take
into account that in the large $t_0$ limit only the terms with
$\omega_{k} + \omega_{j} = \omega_0$ make a remarkable
contribution to the integral in Eq. (\ref{intA}). Therefore the
only essential contribution to this integral comes from the terms
with $ L (t_1)\hat{{\cal A}}_{j,{\vec q}} (t_1) \propto \exp{( \pm
i(\omega_0 -\omega_{j})t_1)}$ (the terms $\propto \exp{(\pm
i(\omega_{j} + \omega_0)t_1)}$ are averaged out at the large $t_0$
limit).  In this case $\,2 \dot{L}(t_1) \, \dot{\hat{{\cal
A}}}_{j} \,(t_1)+ \ddot{L} (t_1)\, \hat{{\cal A}}_{j}\, (t_1)
\propto j^2- k^2$. As a result the factor $j^2- k^2$ in the
equation for $\hat{B}_{k}$ cancels and the $k$-dependence
disappears:
\begin{equation}
\hat{B}\,(t) = 2V \cos{(\omega_0 t)} \hat{Q}; \quad \hat{Q} =
\kappa_0^{-1}L_0^{-1} \sum_{j=1} j \hat{{\cal A}}_{j}. \,\label{Q}
\end{equation}
Here $\hat{Q}$ is the operator of the wave packet of the size
$\sim k_0^{-1}$, where $\,\kappa_0= \omega_0 L_0 /\pi c$.

Eq. (\ref{Q}) for $\hat{B}$ is the key relation of this study: it
accounts for the effect of the oscillations of the optical length
in the case of an infinitely large resonator. This relation
essentially differs from the analogous relation, describing this
effect in the case of a short resonator, which has been studied
earlier \cite{Lobashov,Klimov,Johnston,Saito}: in the latter case
the contribution of only one (or few) so-called "resonant" mode(s)
$j$ into the operator $\hat{B}_k$ is taken into account. However
in the case under consideration there is a continuum of modes
which all are essentially mixed by the oscillations of the optical
length. Therefore, they all contribute to $\hat{B}$. This
circumstance becomes especially clear if one considers the
effective Hamiltonian
\begin{equation}
\hat{H} =\frac{1}{2}\sum_k (\dot{\hat{A}}^2_k + \omega_k^2
\hat{A}^2_k) - V \omega_0 L_0 \cos{(\omega_0 t)}\, \hat{Q}^2
\label{Ham0}
\end{equation}
which corresponds to the equation of motion (\ref{intA}) and the
given $\hat{B}$. Here the last term describes the effective
interaction between  the modes arising from the oscillations of
the optical length. One can see that this interaction is
factorized; all modes contribute to the factors of this
interaction. We note that Hamiltonian (\ref{Ham0}) is analogous to
the one describing the two-phonon decay of a local mode in a
crystal. This allows one to apply the method proposed in
\cite{Hizhrev,EPJ} for a nonperturbative description of this
decay.

To find the number of generated photons, one may diagonalize the
Hamiltonian (\ref{Ham0}) (it is done in \cite{Hizhrev}). However
one can find $N_k$ directly from equations (\ref{intA}), (\ref{Q})
and (\ref{Dasymp}):
\begin{equation}
N_k \simeq \frac{ V^2\omega_k}{2\hbar} \int \!\!\!\!\!\int_0^{t_0}
\!\!\!\! \! dt_1 dt'_1 e^{i(\omega_0 - \omega_k) (t_1 -t'_1)}
D(t_1, t'_1).  \label{N1}
\end{equation}
Here $D(t,t') = \langle 0|\hat{Q} (t_1)  \hat{Q} (t'_1)|0
\rangle$; in the $t_0 \rightarrow \infty$ limit this correlation
function depends on the time difference. The emission rate
$\dot{N}_k = d N_k /d t_0$ now equals
$$
\dot{N}_{k} = (V^2 / 2\hbar) \,\omega_k D(\omega_0 - \omega_{k}),
$$
where $D(\omega)$ is the Fourier transform of the correlation
function $D(t-t')$; here $\omega$ and $\omega_0 - \omega$ are the
frequencies of two emitted photons. Thus, to find the quantum
emission under consideration, one needs to calculate the
correlation function $D(\omega)$. If $V \ll c$ then one can
replace in Eq. (\ref{Q}) for $\hat{Q}$ the field operator
$\hat{{\cal A}}_j$ by $\hat{{\cal A}}_j^{(0)}$. In this
approximation $D(\omega) \approx \hbar \omega /2 \pi^2 c^2
\omega_0^2$ and
$$
\dot{N}(\omega) \approx (v/2 \pi \omega_0)^2 \,\omega ( \omega_0 -
\omega)
$$
($v = V/c$). This equation coincides with that given by the
standard time-dependent perturbation theory. We can see that the
emission under consideration is very  different from the one  in
small resonator: it is stationary, its spectrum is broad, while in
the case of a small resonator, in the resonance conditions, it is
non-stationary and quasi-monochromatic.


To find the emission for an arbitrary $V$, we follow the
calculations given in \cite{EPJ} (see part 2.2.1, the k=2 case).
We use the equation of motion for the operator $\hat{Q}$\,:
\begin{equation}
\hat{Q} (t) = \hat{Q}^{(0)} (t) + 2 v \omega_0 \!\! \int_0^{t}
\!\!\!\!dt_1 G(t-t_1) \cos{(\omega_0 t_1)} \hat{Q} (t_1)
\label{EqQt}
\end{equation}
($t \leq t_0$), which directly follows from Eqs. (\ref{intA}) -
(\ref{Q}). Here $\hat{Q}^{(0)}$ is given by Eq. (\ref{Q}) for
$\hat{Q}$  with $\hat{{\cal A}}_j^{(0)}$ instead of $\hat{{\cal
A}}_j$,
$$
G(t)=( {\rm \Theta} (t)/ \pi \kappa_0^2) \sum_{k=1}^{\kappa_0} k
\sin{\omega_k t}
$$
is the Green function. Using Eq. (\ref{EqQt}) once again (this
time for $\hat{Q} (t_1)$) and inserting it into $D(t,t')$ we find
$$D(t,t') \simeq  d (t,t') + 2v
\omega_0\int_0^{t} \!\!\!\! dt_1 G(t-t_1) \cos{(\omega_0 t_1)}
D_1^*(t_1, t')
$$
\vspace{-7mm}
$$
+v^2 \omega_0^2 \int_0^t  \!\!\!\! dt_1  \!\!\int_0^{t_1}
\!\!\!\!\!\!\! dt_2 G(t-t_1) G(t_1-t_2)e^{-i\omega_0(t_1-t_2)}
D(t_2, t'),
$$
where $d(t,t') = \langle 0|\hat{Q}^{(0)}(t) \hat{Q}(t')|0\rangle$,
and an analogous equation for $d^*(t',t)$. We take into account
only $e^{i\omega_0(t_1-t_2)}$ term of the factor $4 \cos{(\omega_0
t_1)} \cos{(\omega_0 t_2)}$; other terms oscillate fast and drop
out. Above the  term $\propto v$ may be omitted while it also
oscillates fast. Taking into account that in the large time limit
the correlation functions $D(t,t')$ and $d(t,t')$ depend on the
time difference and using the relations
$$
\int_0^{\infty}  \!\!\!\! \!\!dt  \!\! \int_0^t \!\! \!\! dt_1
\!\!\int_0^{t_1} \!\!\!\!\!\!\! dt_2 e^{i\omega t}
=\int_0^{\infty} \!\!\!\!\!\!\! dt_2 e^{i\omega t_2}  \!\!
\int_0^{\infty} \!\!\!\!\!\! d\tau_1 e^{i\omega
\tau_1}\!\!\int_0^{\infty}\!\!\!\!\!\!\!  d\tau e^{i\omega \tau}
$$
\vspace{-5mm}
$$\int_0^{\infty}  \!\!\!\! \!\!dt e^{i\omega t} F(t-t')  \simeq
e^{i\omega t'} \!\!\int_{-\infty}^{\infty}  \!\!\!\! \!\!d\tau
e^{i\omega \tau} F(\tau)= e^{i\omega t'} F(\omega)
$$
($t'\rightarrow \infty, \, \tau = t-t_1, \, \tau_1 = t_1-t_2$),
one gets the following equation for $D(\omega)$
\cite{Hizhrev,EPJ,euro} (the factor $e^{i\omega t'}$ cancels):
$$
D(\omega) \simeq d(\omega)  + v^2 G(\omega) G(\omega-1) D(\omega),
$$
where
$$
G(\omega) = \frac{1}{\pi} \bigg[1+\frac{\omega}{2}
\ln{\frac{|1-\omega|}{|1+\omega|}} \bigg] + \frac{i \omega}{2} \,
\Theta{(1-|\omega|)},
$$
and an analogous equation for $d^*(\omega)$ (we take $\omega_0=1$
for the frequency units). As a result, the number of the emitted
photons with the frequency $\omega$ per unit time and frequency
equals:
\begin{equation}
\dot{N}(\omega) = \frac{(v/2 \pi)^2 \omega(1-\omega)}{|1-v^2 G^*
(\omega) G(1 - \omega) |^2}. \label{arbit}
\end{equation}
This expression describes the emission under consideration for any
value of $v$. If $v \ll 1$, then the intensity of the emission
increases quadratically with $v$. However, if $v \gg1$, then the
dependence on $v$ is the opposite: $\dot{N} \propto v^{-2}$. If
$v$ approaches the value
$$
v_{r} =4 \pi/ \sqrt{\pi^2 + (4- \ln{3})^2} \approx 2.94,
$$
then the resolvent at $\omega =$ 1/2 diverges and the emission is
resonantly enhanced (see Figs. 1 and 2). The given value of $v_r$
is close to $\pi$, i.e. the value of $v$ which corresponds to the
resonance between the oscillations of the optical length and the
generated wave, when the distance between the turning points of
$l$ coincides with the half-wave of the generated photons.
\begin{figure}
\includegraphics[width=12cm]{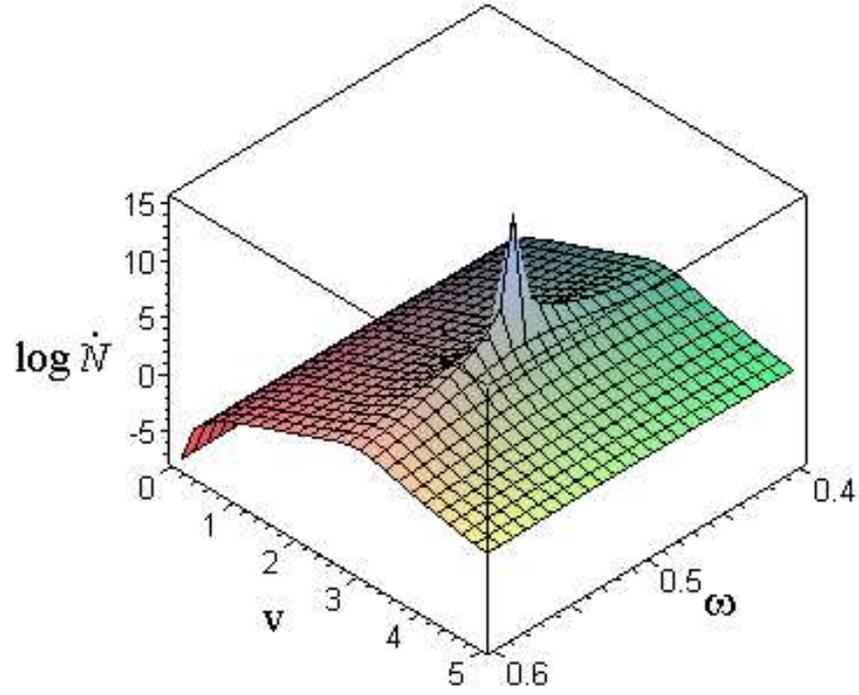}
\caption{The logarithm of the rate of the emission of photons
$\dot{N}$ by a medium with time-dependent refractive index $n =
n_0 + n'_0 \cos{\omega_0 t}$, as a function of the frequency of
emission $\omega$ and maximal velocity $v$ of the optical length
(the units $\omega_0 =1$ and $c=1$ are used).}
\end{figure}
\begin{figure}
\includegraphics[width=12cm]{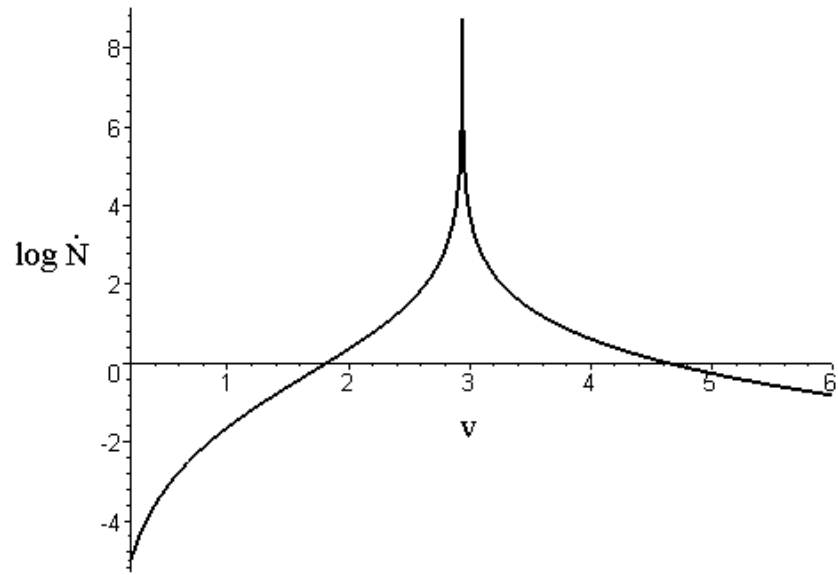}
\caption{The logarithm of the integrated rate of the emission of
photons $\dot{\rm N} = \int \!\! \dot{N} (\omega)d \omega $ by a
medium with time-dependent refracting index, as a function of the
maximal velocity $v$ of the optical length (the same units as in
Fig. 1 are used).}
\end{figure}

To estimate the intensity of a laser beam which can give $v  =
n'_0 \omega_0 L^{(0)}/c \sim 1$ and which can cause the resonant
enhancement of the two-photon emission, we  take $n'_0 \sim
n^{(2)} I$, where $I$ is the intensity of the laser light,
$n^{(2)} \sim 10^{-15} cm^2/W$, - a typical non-resonant value of
$n^{(2)}$ in crystals \cite{Shen}. We also take $\omega_0
L^{(0)}/c \sim 10^{5}$and $n^{(2)} \sim 10^{-15} cm^2/W$. We get
$I \sim 10^{10} W/cm^2$. Such intensity of the laser light is
experimentally achievable.

In our consideration we describe  the emission in the direction
$x$, which was chosen arbitrarily. This means that photons are
emitted in all directions. The positive and the negative direction
along the $x$ axis are not distinguished in our case. Therefore
the pairs of photons with the wave vectors $\pm {\vec q}$ and $\pm
\alpha {\vec q}$ along the $x$ axis are equally generated by the
medium (here $\alpha = \omega_0 / \omega -1$).

Only a spontaneous emission was considered above. If $N_{\vec q}$
photons with the wave-vector ${\vec q}$ and  the frequency $\omega
= c q < \omega_0$ are fall into a medium with $n$ oscillating in
time, then also stimulated processes give a contribution to the
emission. As a result, an additional factor $1+ N_{\vec q}$
appears in the equation for the intensity of the emission of
photons with the wave-vectors ${\vec q}$ and ${\vec q} = - \alpha
{\vec q}$, where $\alpha = \omega_0 / \omega -1$. It is essential
to underline that the presence of photons with the wave-vector
${\vec q}$ leads to an enhancement of the emission of photons not
only with the same wave vector but also of photons with the
wave-vector $-\alpha {\vec q}$. E.g. the photons with the wave
number $q=\omega_0/2c$ do stimulate the emission of photons with
the wave vector $= -{\vec q}$. This is a well-known fact
\cite{Pepper}: a medium with a refractive index oscillating in
time performs the phase-conjugated reflection of photons with half
a frequency. In our case, in the resonance condition it takes
place with a high efficiency.

A medium with oscillating refractive index is not the only
physical system where the optical length oscillates in time; a
resonator with  vibrating mirror(s) gives another example. In the
latter case the maximal velocity of the oscillations of the
mirror(s) is limited by the velocity of light. However a
reflecting border in a dielectric medium can be put to oscillate
with $V>c$, e.g. by means of a strong laser beam which
periodically changes its direction so that the position of the
area, being illuminated by the beam, moves forth and back with
$V>c$. In this case one can get strong enhancement of the
two-photon emission if $V$ approaches $2.9 c$.  Note also that in
a small resonator (cavity) with vibrating walls one can also get
strong enhancement of the generation of photons of a mode if it is
in parametric resonance with the oscillations (see, e.g.
\cite{Casimir,Law,More,Janovich,Mundarain,Dodonov,Jenson,Visser,
Jacobs,Rzazewski,Schaller}).

Finally, we note that the emission under consideration can be
generated by any strong coherent long-wave excitation, which
periodically modulates $n$. A similar emission of other bosons as
well as the resonant enhancement of this emission is also possible
in a periodically time-dependent medium, analogously to the
above-described resonant enhancement of the two-photon emission.
To prove the aforementioned  we consider the quantum field, which
satisfies the time-dependent Klein-Gordon equation
$$
[\partial^2/ \partial t^2 -c^2 \hat{\nabla}^2 + m^2 ] \hat{A}
=-\partial^2 \eta(t) \hat{A}/ \partial t^2.
$$
In this case the above presented consideration of the two-quantum
emission holds also if one replaces the frequency of a photon
$\omega = cq$ by the frequency of the particle $\sqrt{c^2q^2 +
m^2}$. The number of the emitted quanta (particles and
antiparticles) in the time unit is also described by the Eq.
(\ref{arbit}) if $G(\omega)$ is replaced by the difference
$G(\omega)- G_1(\omega)$ where $G_1(\omega)$ can be obtained from
$G(\omega)$ by replacing $1 -\omega$ by $2 m -\omega$; the
two-particle emission under consideration exists only if the rest
mass of the particle-antiparticle pair $2 m $ (in the
$\omega_0=\hbar =c=1$ units) does not exceed $1$. This result may
be of interest to the physics of condensed matter, e.g. to the
generation of phonon pairs in a semiconductor by a strong
microwave or to a two-phonon decay of the strong phonon wave
generated in CARS experiments. It may also offer interest for the
astrophysics as a possible mechanism of a powerful emission of
particles.

To sum up, a solution of the problem of the two-quantum emission
of a periodically time-dependent medium has been given. It has
been found that, if the maximal value of the velocity of the
oscillating optical length approaches the critical value 2.94$c$,
then a strong enhancement of the two-photon emission takes place.
It has also been found that a medium with the resonant
oscillations of the refractive index may carry out the
phase-conjugated reflection with high efficiency. A similar
resonant enhancement of other types of two-quantum emission has
also been predicted.

Acknowledgement. The research was supported by the ETF Grant No
5023 and by the US NRC Twinning Program.
%

\end{document}